\documentclass[aps,twocolumn]{revtex4}
\usepackage{amsmath,amssymb}

\usepackage{epsfig}
\usepackage{graphicx}

\newcommand{\be}{\begin{equation}}
\newcommand{\ee}{\end{equation}}
\newcommand{\bea}{\begin{eqnarray}}
\newcommand{\eea}{\end{eqnarray}}
\newcommand{\nn}{\nonumber}

\begin{document}

\title{ On braneworld cosmologies from six dimensions, and absence thereof}

\author{Georgios Kofinas\footnote{gkofin@phys.uoa.gr; kofinas@ffn.ub.es}}

\date{\today}

%\address{~}

\address{Departament de F{\'\i}sica Fonamental,
Universitat de Barcelona, Diagonal 647, 08028 Barcelona, Spain}

\begin{abstract}

We consider (thin) braneworlds with conical singularities in
six-dimensional Einstein-Gauss-Bonnet gravity with a bulk
cosmological constant. The Gauss-Bonnet term is necessary in six
dimensions for including non-trivial brane matter. We show that this
model for axially symmetric bulks does not possess isotropic
braneworld cosmological solutions.

\end{abstract}

\maketitle

Much work on braneworlds in six-dimensional spacetimes has been
done, especially during the last two years. In classical
six-dimensional gravity \cite{gra} or supergravity \cite{sugra}
theories a codimension-two object induces a conical singularity
\cite{coni}, and a cancelation occurring between the brane tension
and the bulk gravitational degrees of freedom gives rise to a
vanishing effective cosmological constant. Couplings of
six-dimensional gravity to sigma models have been discussed in
\cite{sigma}. Other works have focused on static/time-dependent
solutions and issues of stability \cite{static}. It is known that
six-dimensional Einstein gravity cannot support a (thin gravitating)
braneworld with a non-trivial matter content different than a brane
tension \cite{cline}. Proposals for generalizing the brane equation
of state, or deriving cosmologies have been made \cite{propo}. The
situation can be improved if a Gauss-Bonnet term is added to the
bulk action, in which case the generic matching conditions of a
3-brane with conical singularities were derived in \cite{ruth} (see
also \cite{charmousis, others, efforts}). The conservation equation
of the braneworld was derived in \cite{kofinas}. In the present
paper we consider the isotropic braneworld cosmology of this theory
for axially symmetric bulks around the defect, and with a bulk
cosmological constant. We show that the model is incompatible with
such braneworld configurations.

\par
We consider the total gravitational brane-bulk action
 \bea &&
\!\!\!\!\!\!\!S_{gr}=\frac{1}{2\kappa_{6}^{2}}\!\int
\!d^6x\sqrt{\!-|\verb"g"|}
\left\{\mathcal{R}-2\Lambda_{6}+\alpha\,\Big(\mathcal{R}^{2}\!-4\mathcal{R}_{AB}\,
\mathcal{R}^{AB} \right.\nn
\\
&&\!\!\!\!\!\!\left.~{}+\mathcal{R}_{ABCD}\,\mathcal{R}^{ABCD}\Big)\!\right\}+
\frac{r_{c}^{2}}{2\kappa_{6}^{2}}\!\int \!d^4x\sqrt{\!-|g|}
\left(R-2\Lambda_{4}\right)\!, \label{6action}\eea where
calligraphic quantities refer to the bulk metric tensor $\verb"g"$,
while the regular ones to the brane metric tensor $g$. The
Gauss-Bonnet coupling $\alpha$ has dimensions $(length)^{2}$ and is
defined as
 \bea
\alpha=\frac{1}{ 8g_{s}^{2}}\,,
 \eea
with $g_{s}$ the string energy scale, while from the induced-gravity
crossover lenght scale $r_{c}$ we can define
 \bea
r_{c}=\frac{\kappa_6}{\kappa_4}=\frac{M_4}{M_6^2}\,.
 \label{distancescale}
 \eea
Here, $M_6$ is the fundamental six-dimensional Planck mass
$M_{6}^{-4}\!=\!\kappa_{6}^{2}\!=\!8\pi G_{6}$, while $M_{4}$ is
given by $M_{4}^{-2}\!=\!\kappa_{4}^{2}\!=\!8\pi G_{4}$. The brane
tension is
 \bea
\lambda={\Lambda_4 \over \kappa_4^2}\,.
 \eea
The field equations arising from the action (\ref{6action}) are \bea
&&\!\!\!\!\!\!\!\mathcal{G}_{AB}-\frac{\alpha}{2}(\mathcal{R}^2-4\mathcal{R}_{CD}\mathcal{R}^{CD}+\mathcal{R}_{CDEF}\mathcal{R}
^{CDEF})\verb"g"_{AB}+2\alpha\nn\\
&&\!\!\!\!\!\!\!\times(\mathcal{R}\mathcal{R}_{AB}\!-\!2\mathcal{R}_{AC}\mathcal{R}_{B}^{\,\,\,\,C}\!\!-\!
2\mathcal{R}_{ACBD}\mathcal{R}^{CD}\!\!+\!\mathcal{R}_{ACDE}\mathcal{R}_{B}^{\,\,\,\,CDE})\nn\\
&&\,\,\,\,\,\,\,\,\,\,\,\,\,\,\,\,\,\,\,
\,\,\,\,\,\,\,\,\,\,\,\,\,\,\,=\kappa_{6}^2
\mathcal{T}_{AB}-\Lambda_{6}
\verb"g"_{AB}+\kappa_{6}^2\,^{(loc)}\!T_{AB}\,\delta^{(2)},\label{6eqs}
\eea where $\mathcal{T}_{AB}$ is a regular bulk energy-momentum
tensor, $T_{AB}$ is the brane energy-momentum tensor,
$^{(loc)}\!T_{AB}=T_{AB}-\lambda
g_{AB}-(r_{c}^{2}/\kappa_{6}^2)G_{AB}$, and $\delta^{(2)}$ is the
two-dimensional delta function. Capital indices $A,B,...$ are
six-dimensional. Assuming that the bulk metric in the brane-adapted
coordinate system takes the axially symmetric form
 \bea
ds_{6}^2=dr^2+L^2(x,r)d\varphi^2+g_{\mu\nu}(x,r)dx^{\mu}dx^{\nu},\label{6metric}
\eea with $g_{\mu\nu}(x,0)$ being the braneworld metric and
$\varphi$ having the standard periodicity $2\pi$, under the usual
assumptions for conical singularities
$L(x,r)=\beta(x)r+\mathcal{O}(r^2)$ for $r\approx 0$,
$\partial_{r}L(x,0)=1$, $\partial_{r}g_{\mu\nu}(x,0)=0$, the general
matching conditions for imbedding the 3-brane in the six-dimensional
theory (\ref{6action}) were found in \cite{ruth} (see also
\cite{charmousis}) as follows \bea &&\!\!\!\!\!\!
K^{\alpha\lambda}_{\,\,\,\,\,\,\,\lambda} K_{\alpha\mu\nu}\!-\!
K^{\alpha\lambda}_{\,\,\,\,\,\,\,\mu}K_{\alpha\nu\lambda}
\!+\!\frac{1}{2}(
K^{\alpha\lambda\sigma}K_{\alpha\lambda\sigma}\!-\!
K^{\alpha\lambda}_{\,\,\,\,\,\,\,\lambda}K_{\alpha\,\,\,\sigma}^{\,\,\,\sigma})g_{\mu\nu}\nn\\
&&\!\!\!\!\!\!\!+\Big(\!\beta^{-1}\!\!-\!1\!+\!\frac{r_{c}^2}{8\pi\alpha\beta}\!\Big)G_{\mu\nu}\!+\!
\frac{\kappa_{6}^{2}\lambda\!-\!2\pi(1\!-\!\beta)}{8\pi\alpha\beta}g_{\mu\nu}\!=\!
\frac{\kappa_{6}^{2}}{8\pi\alpha\beta}T_{\mu\nu}.
\nn\\
\label{matching} \eea  Here,
$K_{\alpha\mu\nu}=\verb"g"(\nabla_{\mu}n_{\alpha},\partial_{\nu})=n_{\alpha\mu;\nu}$
(at $r\!=\!0^{+}$) denote the extrinsic curvatures of the brane
(symmetric in $\mu,\nu$), where $n_{\alpha}$ ($\alpha=1,2$) are
arbitrary unit normals to the brane (indices $\alpha, \beta,...$ are
lowered/raised with the matrix
$\verb"g"_{\alpha\beta}=\verb"g"(n_{\alpha},n_{\beta})$ and its
inverse $\verb"g"^{\alpha\beta}$), while $\nabla$ (also denoted by
$;$) refers to the Christoffel connection of $\verb"g"$. For
extracting this singular part of equations (\ref{6eqs}), one has to
focus on the worst behaving pieces with the structure
$\delta(r)/L\sim \delta(r)/r$. Note that with respect to local
rotations $n_{\alpha}\rightarrow
O^{\,\,\,\beta}_{\alpha}(x^{A})\,n_{\beta}$,
$K_{\alpha\mu\nu}\rightarrow
O^{\,\,\,\beta}_{\alpha}K_{\beta\mu\nu}$ transforming as a vector,
thus Eq.(\ref{matching}) is invariant under changes of the normal
frame.

Focusing on the $\mathcal{O}(1/r)$ terms in the $r\mu$ components of
equations (\ref{6eqs}) (which cannot be canceled by any regular
$\mathcal{T}_{AB}$ in (\ref{6eqs})) we obtain the equation \bea
&&\!\!\!\!\!\!\!
\mathcal{R}^{\!\alpha\sigma}_{\,\,\,\,\,\,\nu\sigma}K_{\alpha\,\,\,\lambda}^{\,\,\,\lambda}\!-\!
\mathcal{R}^{\!\alpha\sigma}_{\,\,\,\,\,\lambda\sigma}K_{\!\alpha\,\,\,\nu}^{\,\,\lambda}\!-\!
\mathcal{R}^{\!\alpha\lambda}_{\,\,\,\,\,\,\nu\sigma}K_{\!\alpha\,\,\,\lambda}^{\,\,\sigma}\!=
\frac{\beta_{,\mu}}{\beta}\!\Big[G_{\nu}^{\mu}\!-\!\frac{1}{4\alpha}\delta_{\nu}^{\mu}\nn\\
&&\!\!\!\!\!\!\!+K^{\alpha\sigma}_{\,\,\,\,\,\,\,\nu}
K_{\!\alpha\,\,\,\sigma}^{\,\,\mu}\!-\!
K^{\alpha\sigma}_{\,\,\,\,\,\,\,\sigma}K_{\alpha\,\,\,\nu}^{\,\,\,\mu}
\!+\!\frac{1}{2}\!(
K^{\alpha\sigma}_{\,\,\,\,\,\,\,\sigma}K_{\!\alpha\,\,\lambda}^{\,\,\lambda}
\!-\!K^{\alpha\sigma\lambda}\!K_{\alpha\sigma\lambda}\!)\delta_{\nu}^{\mu}\!\Big]\!.\nn\\
\label{christmas} \eea In \cite{kofinas} it was shown that equation
(\ref{christmas}) is equivalent to the standard conservation
equation on the brane \bea T^{\mu}_{\nu|\mu}=0, \label{conservation}
\eea where $|$ refers to the Christoffel connection
$\gamma_{\mu\nu\lambda}\!=\!\verb"g"(\nabla_{\lambda}\partial_{\nu},\partial_{\mu})$
of the induced brane metric $g_{\mu\nu}$. Thus, we do not consider
equation (\ref{christmas}) further, but only equation
(\ref{conservation}).

From the $\mathcal{O}(1/r)$ part of the $rr$ component of equations
(\ref{6eqs}) we obtain the following equation, valid at the position
of the brane \bea &&\!\!\!\!\!\!\!\!\!\!\!\!\!\!
g^{\mu\nu}g_{\mu\nu}^{\,\prime}[\,4R\!-\!(g^{\kappa\lambda}g_{\kappa\lambda}^{\,\prime})^2\!-\!
3g^{\kappa\lambda\,\prime}\,
g_{\kappa\lambda}^{\,\prime}\!+\!2\alpha^{-1}]\!-\!
8R^{\mu\nu}g_{\mu\nu}^{\,\prime}\nn\\
&&\,\,\,\,\,\,\,\,\,\,\,\,\,\,\,\,\,\,\,\,\,\,\,\,\,\,\,\,\,\,\,\,\,\,\,\,\,\,\,\,\,\,\,\,\,\,\,\,\,
\,\,\,\,\,\,\,\,\,\, -2g_{\mu\nu}^{\,\prime} g^{\mu\kappa\,\prime}
g^{\nu\lambda\,\prime}g_{\kappa\lambda}\!=\!0, \label{rr} \eea where
a prime denotes differentiation with respect to $r$. Note that in
the coordinates (\ref{6metric}) it is
$K_{r\mu\nu}\!=\!g_{\mu\nu}^{\,\prime}/2$, $K_{\varphi\mu\nu}=0$. We
will transform equation (\ref{rr}) to an equivalent and simpler
form. To do so, we contract the matching conditions (\ref{matching})
with $g^{\mu\nu\,\prime}$ and replace from this equation the last
term of equation (\ref{rr}). Making also use of the trace of
equations (\ref{matching}), equation (\ref{rr}) gets the form \bea
(\sigma_{1}G^{\mu\nu}\!+\!\sigma_{2}g^{\mu\nu}\!+\!\sigma_{3}T^{\mu\nu})g_{\mu\nu}^{\,\prime}=0,
\label{easy} \eea where \bea
\sigma_{1}=1\!+\!\frac{r_{c}^{2}}{8\pi\alpha}\,\,,\,\,
\sigma_{2}=\frac{\kappa_{6}^{2}\lambda\!-\!2\pi}{8\pi\alpha}\,\,,\,\,
\sigma_{3}=-\frac{\kappa_{6}^{2}}{8\pi\alpha}. \label{constants}
\eea This equation is linear and homogeneous in the components of
the extrinsic curvature, does not contain the deficit angle $\beta$,
and will facilitate our analysis.

The only nontrivial remaining components of equations (\ref{6eqs})
with a $\mathcal{O}(1/r)$ part are the $\mu\nu$ ones, which give the
equation \bea
&&\!\!\!\!\!\!4\frac{\beta_{,\kappa}}{\beta}g^{\kappa\lambda}[\mathcal{R}_{r(\mu|\lambda|\nu)}\!\!-\!\!
\mathcal{R}_{r\sigma\tau(\mu}\,g_{\nu)\lambda}\,g^{\sigma\tau}\!\!-\!\mathcal{R}_{r\sigma\lambda\tau}
g^{\sigma\tau}\!g_{\mu\nu}]\!=\!\textsf{c}\,G_{\!\mu\nu}\nn\\
&&\!\!\!\!\!\!+\frac{5}{4}g_{\mu\kappa}^{\,\prime}g_{\nu\lambda}^{\,\prime}g^{\kappa\lambda\,\prime}
\!\!+\!\!\Big(\!\!4R\!-\!5\textsf{b}\!-\!3\textsf{c}^{2}\!\!+\!\frac{2}{\alpha}\!\Big)
\frac{g_{\mu\nu}^{\,\prime}}{8}\!+\!\textsf{c}\Big(\!5\textsf{b}\!+\!\textsf{c}^{2}\!\!-\!\frac{2}{\alpha}\!\Big)
\frac{g_{\mu\nu}}{8}\nn\\
&&\!\!\!\!\!\!-\!2R^{\lambda}_{(\mu}g_{\nu)\lambda}^{\,\prime}\!\!+\!R^{\kappa\lambda}
g_{\kappa\lambda}^{\,\prime}g_{\mu\nu}\!\!+\!R_{\mu\kappa\nu\lambda}g^{\kappa\lambda\,\prime}\!\!\!-
\!\frac{1}{2}g_{\kappa\sigma}^{\,\prime}g_{\lambda\rho}^{\,\prime}g^{\kappa\lambda\,\prime}g^{\sigma\rho}
g_{\mu\nu}\nn\\
&&\!\!\!\!\!\!\!+
\textsf{c}g_{\mu\kappa}^{\,\prime}g_{\nu\lambda}^{\,\prime}g^{\kappa\lambda}
\!\!+\!\frac{1}{2}[2\,\widehat{g^{\,\prime\prime}}_{\!\!\!\!\!\kappa(\mu}\,g_{\nu)\lambda}^
{\,\prime}g^{\kappa\lambda}\!+\!\!(\hat{\texttt{f}}\!+\!\textsf{c}\,\hat{\textsf{f}}\,)g_{\mu\nu}
\!\!-\!\textsf{c}\,\widehat{g^{\,\prime\prime}}_{\!\!\!\!\!\mu\nu}\!-\!\hat{\textsf{f}}g_{\mu\nu}^{\,\prime}]\nn\\
&&\!\!\!\!\!\!+\frac{\widehat{L''}}{2\beta}\!\Big[\!4G_{\mu\nu}\!-\!\textsf{c}g_{\mu\nu}^{\,\prime}\!+\!
g_{\mu\kappa}^{\,\prime}g_{\nu\lambda}^{\,\prime}g^{\kappa\lambda}\!+\!\Big(\!\frac{\textsf{b}\!+\!\textsf{c}^{2}}{2}\!-
\!\frac{1}{\alpha}\!\Big)g_{\mu\nu}\!\Big], \label{diffi} \eea where
for abbreviating the expression we have defined \bea
\!\!\textsf{b}\!=\!g^{\mu\nu\,\prime}\,g_{\mu\nu}^{\,\prime}\,\,,\,\,\textsf{c}\!=\!g^{\mu\nu}g_{\mu\nu}^{\,\prime}\,\,,\,\,
\texttt{f}\!=\!g^{\mu\nu\,\prime}\,g_{\mu\nu}^{\,\prime\prime}\,\,,\,\,\textsf{f}\!=\!
g^{\mu\nu}g_{\mu\nu}^{\,\prime\prime}, \label{detail} \eea and an
overhat means the regular part of the corresponding quantity.

The only equations remaining to be valid on the brane come from the
regular part of the system (\ref{6eqs}).

There are two cases concerning the form of the possible braneworld
solutions: (a) $K_{\alpha\mu\nu}$ is not identically zero, and (b)
$K_{\alpha\mu\nu}=0$. In the case (a) one has to consider all the
previous equations together. In the case (b) the matching condition
(\ref{matching}) takes the form of purely 4-dimensional Einstein
gravity, equation (\ref{christmas}) implies $\beta\!=\!$ constant,
equations (\ref{rr}), (\ref{easy}) are identically satisfied, while
equation (\ref{diffi}) implies $\widehat{L''}\!=\!0$. Considering
the six-dimensional Ricci scalar, this contains singular
$\delta(r)/r$ terms, and, in general, also terms of the form $1/r$
(multiplied by $g_{\mu\nu}^{\,\prime}$). Thus, in the case (b) these
last $1/r$ terms vanish, while in case (a) tidal forces appear in
the vicinity of the braneworld. Our aim is to find any 4-dimensional
isotropic cosmology compatible with the model or to show that no
such cosmology exists. We are interested here in a bulk with a pure
cosmological constant $\Lambda_{6}$; however, for possible use of
the present formulation elsewhere we let $\mathcal{T}_{AB}$
non-vanishing.

We consider the bulk cosmological metric of the form (\ref{6metric})
\bea &&\!\!\!\!\!\!
ds_{6}^2\!=\!dr^{2}\!+\!L^{2}(t,r\!)d\varphi^{2}\!-\!n^{2}(t,r\!)dt^{2}\!+\!a^{2}(t,r\!)
\gamma_{ij}(x)dx^{i}\!dx^{j},\nn\\
&&\label{cosmobulk} \eea where $\gamma_{ij}$ is a maximally
symmetric 3-dimensional metric characterized by its spatial
curvature $k=-1,0,1$. For the metric (\ref{cosmobulk}) the matching
conditions (\ref{matching}) are written equivalently as \bea &&
\!\!\!\!\!\!\!\!\!\!\!\!\!\!\!A^{2}=\Big(\!1\!-\!\frac{1}{\beta}\!\Big)\Big(\!X\!+\!
\frac{1}{12\alpha}\!\Big)\!+\!\frac{\sigma_{3}}{3\beta}\,^{(loc)}\!T^{t}_{t}\label{roa}\\
&& \!\!\!\!\!\!\!\!\!\!\!\!\!\!\!\!\!\!
AN=\Big(\!1\!-\!\frac{1}{\beta}\!\Big)\Big(\!Y\!+\!
\frac{1}{12\alpha}\!\Big)\!+\!\frac{\sigma_{3}}{6\beta}(^{(loc)}\!T^{\mu}_{\mu}\!-\!2\,^{(loc)}\!T^{t}_{t}),
\label{rua} \eea where
\bea && \,\,\,\,\,\,A=\frac{a'}{a}\,\,\,\,\,,\,\,\,\,\,N=\frac{n'}{n}\\
&&\!\!\!\!\!\!\!\!\!X=H^{2}\!+\!\frac{k}{a^{2}}\,\,\,\,\,,\,\,\,\,\,Y=\frac{\dot{H}}{n}\!+\!H^{2},
\label{lake} \eea with $H\!=\!\dot{a}/na$ being the Hubble parameter
of the brane and a dot denotes differentiation with respect to $t$.
Throughout, the lapse function $n$ is left undetermined and does not
affect the analysis since it corresponds to the temporal gauge
choice on the brane. The matter on the brane is taken to be a
perfect fluid with energy density $\rho$ and pressure $p\!=\!w\rho$.
Equation (\ref{easy}) takes the simple form \bea N=fA, \label{es}
\eea where \bea
f=3\frac{\sigma_{3}p\!+\!\sigma_{2}\!-\!\sigma_{1}(X\!+\!2Y)}
{\sigma_{3}\rho\!-\!\sigma_{2}\!+\!3\sigma_{1}X}. \label{f} \eea The
$tt$ component of equation (\ref{diffi}) is \bea
&&\!\!\!\!\!\!\!\!A\Big(\!A^{2}\!-\!X\!-\!\frac{1}{4\alpha}\!+\!\frac{2\widehat{a''}}{a}\Big)\!+\!
\frac{\widehat{L''}}{\beta}\Big(\!A^{2}\!-\!X\!-\!\frac{1}{12\alpha}\!\Big)\!=\!0,
\label{hop} \eea while the $ij$ components of the same equation give
\bea &&\!\!\!\!\!\!\!\!\!\!\!\frac{4\dot{\beta}}{n
\beta}\Big[\!\frac{\dot{A}}{n}\!+\!H(A\!-\!N)\!\Big]
=NX\!+\!2AY\!-\!3NA^{2}\!+\!\frac{N\!+\!2A}{4\alpha}\nn\\
&&\!\!\!\!\!\!\!\!\!\!\!\!\!\!\!\!-2(A\!+\!N)\frac{\widehat{a''}}{a}
\!-\!2A\frac{\widehat{n''}}{n}\!+\!\frac{\widehat{L''}}{\beta}
\!\Big[\!X\!+\!2Y\!\!-\!\!A(A\!+\!2N)\!+\!\frac{1}{4\alpha}\!\Big]\!.
\label{tria} \eea

From equations (\ref{roa}), (\ref{rua}), (\ref{es}), we can find the
extrinsic curvature and the deficit angle \bea
&&\!\!\!\!\!\!\!\!\!\!\!\!\!\!
A^{2}\!=\!\frac{2(\sigma_{3}\rho\!-\!\sigma_{2}\!-\!\frac{\sigma_{1}}{4\alpha})
(Y\!-\!X)\!+\!3\sigma_{3}(\rho\!+\!p)(X\!+\!\frac{1}{12\alpha})}{\sigma_{3}
(\rho\!+\!9p)\!+\!8\sigma_{2}\!-\!6\sigma_{1}(X\!+\!3Y)},
\label{ext} \eea
 \bea
&&\!\!\!\!\!\!\!\!\!\!\!\!
\beta=\frac{\sigma_{3}\rho\!-\!\sigma_{2}\!+\!3\sigma_{1}X}{3(X\!-\!A^{2}\!+\!\frac{1}{12\alpha})}\,.
\label{deficit} \eea

The regular part of the $r\mu$ components of equations (\ref{6eqs})
gives on the brane \bea
&&\!\!\!\!\!\!\!\!\Big(\!X\!-\!A^{2}\!+\!\frac{1}{4\alpha}\!+\!
2H\frac{\dot{\beta}}{n\beta}\!\Big)\frac{\dot{A}}{nA}
\!+\!H\Big(\!1\!-\!\frac{N}{A}\!\Big)\Big(\!X\!-\!A^{2}\!+\!\frac{1}{4\alpha}\!\Big)\nn\\
&&\!\!\!\!\!\!\!+
\Big[\!2H^{2}\Big(\!1\!-\!\frac{N}{A}\!\Big)\!-\!\frac{N}{A}\Big(\!X\!-\!A^{2}\!+\!\frac{1}{12\alpha}\!\Big)\!\Big]
\frac{\dot{\beta}}{n\beta}=\frac{n\kappa_{6}^{2}\,\mathcal{T}^{t}_{r}}{12\alpha
A}. \label{mammut} \eea (Note that for the case (b) equation
(\ref{mammut}) is trivially satisfied with
$\mathcal{T}^{t}_{r}\!=\!0$). Similarly, the regular part of the
$rr$ component of equations (\ref{6eqs}) gives \bea
&&\!\!\!\!\!\!\!\!
\Big(\!X\!\!-\!\!A^{2}\!+\!\frac{1}{4\alpha}\!+\!2Y\!\!-\!2AN\!\Big)\!\frac{H\dot{\beta}}{n\beta}\!+\!
\Big(\!\!X\!\!-\!\!A^{2}\!+\!\frac{1}{4\alpha}\!\Big)\!\Big(\!Y\!\!-\!\!AN\!+\!\frac{1}{4\alpha}\!\Big)\nn\\
&&\!\!\!\!\!\!\!+\Big(\!X\!\!-\!\!A^{2}\!+\!\frac{1}{12\alpha}\!\Big)\!
\Big[\frac{1}{n}\!\Big(\!\frac{\dot{\beta}}{n\beta}\!\Big)^{^{\!\!.}}\!\!+
\!\Big(\!\frac{\dot{\beta}}{n\beta}\!\Big)^{\!\!2}\Big]\!=\!\frac{\Lambda_{6}\!-\!\kappa_{6}^{2}\mathcal{T}^{r}_{r}}
{12\alpha}\!+\!\frac{1}{16\alpha^{\!2}}. \label{hope} \eea The other
regular parts of the system (\ref{6eqs}) (namely, equations
$\varphi\varphi$, $\mu\nu$) contain the quantities $\widehat{a''}$,
$\widehat{n''}$. Considering, now, the bulk system (\ref{6eqs}), it
is expected, due to the Bianchi-Bach-Lanczos identities, that one of
these equations, say the $ij$ one, is redundant and it is derived
from the other equations of the system. Thus, both equations
(\ref{tria}), and the $ij$ regular part of (\ref{6eqs}) are
redundant. The remaining two regular equations $\varphi\varphi$,
$tt$ determine $\widehat{a''}$, $\widehat{n''}$, while equation
(\ref{hop}) gives the value of $\widehat{L''}$. Equation
(\ref{mammut}), when $A, \beta$ are substituted from (\ref{ext}),
(\ref{deficit}) becomes an equation for $\dot{Y}$ (i.e. $\ddot{H}$,
or more precisely an autonomous equation for $\dddot{a}$) which is
the candidate cosmological equation of the model. This equation
remains to be compatible with equation (\ref{hope}), which means
that the compatibility has to be checked at the order $\ddot{Y}$.
For the case (b), equation (\ref{hope}) becomes \bea
\Big(\!X\!+\!\frac{1}{4\alpha}\!\Big)\!\Big(\!Y\!+\!\frac{1}{4\alpha}\!\Big)\!=\!
\frac{\Lambda_{6}\!-\!\kappa_{6}^{2}\mathcal{T}^{r}_{r}}{12\alpha}
\!+\!\frac{1}{16\alpha^{2}}, \label{caseb} \eea which is seen to be
inconsistent with the solution
$X\!=\!(\beta\!-\!\sigma_{1})^{-1}(\sigma_{3}\rho\!-\!\sigma_{2}\!-\!\beta/4\alpha)/3$
of the matching conditions (\ref{matching}).

Continuing with the general case (a), we define the variables \bea
&&\!\!\!\!\!\!\!\!\!\!\!\!\!\!\!\!x=X\!+\!\frac{1}{12\alpha}\,\,\,\,,\,\,\,\,
P=\sigma_{3}\rho\!-\!\sigma_{2}\!+\!3\sigma_{1}X\,\,\,\,,\,\,\,\,
{\ss}=\frac{1}{\beta}\,, \label{variables} \eea and replacing
$\dot{A}$ from equation (\ref{roa}), we write the system of
equations (\ref{mammut}), (\ref{hope}) equivalently as
\begin{eqnarray} &&\!\!\!\!\!\!\!\!\!\!\!
\Big(\!x\!-\!\frac{1}{12\alpha}\!-\!\frac{k}{a^{2}}\!\Big)\Big(\frac{d\ln\!{\ss}}{d\ln
\!a}\Big)^{\!2}\!-\!\frac{1}{6}
\Big(\!{\ss}P\!-\!6fA^{2}\!+\!\frac{1}{2\alpha}\!\Big)\frac{d\ln\!{\ss}}{d\ln
\!a}\nn\\
&&\,\,\,\,\,\,\,\,\,\,\,\,\,\,\,\,\,\,\,\,\,\,\,\,\,\,\,\,\,\,\,\,\,\,\,\,\,\,\,\,\,\,\,\,\,
\,\,\,\,\,\,\,\,\,\,\,\,\,\,\,\,\,\,\,\,\,\,\,\,\,\,\,\,\,\,\,\,\,\,\,\,\,\,\,\,\,\,\,\,\,
=\frac{n\kappa_{6}^{2}A\mathcal{T}^{t}_{r}}{4\alpha H {\ss}P}
\label{viki}\\
&&\!\!\!\!\!\!\!\!\!\!{\ss}P\Big(\!x\!-\!\frac{1}{12\alpha}\!-\!\frac{k}{a^{2}}\!\Big)\frac{d^{2}\!\ln\!{\ss}}{d(\ln
\!a\!)^{2}}\!-\! {\ss}P\Big[\frac{{\ss}P}{6}(2\!+\!5f)\!-\!(1\!+\!f)\frac{k}{a^{2}}\nn\\
&&\!\!\!\!\!\!\!\!\!+\Big(\!x\!+\!\frac{1}{12\alpha}\!\Big)(1\!-\!\!f)\!\Big]\frac{d\ln\!{\ss}}{d\ln
\!a}\!-\!\frac{1}{6}\Big(\!{\ss}P\!+\!\frac{1}{2\alpha}\!\Big)\Big[\frac{1}{\alpha}\!-\!{\ss}P(1\!+\!f)\!\Big]\nn\\
&&\,\,\,\,\,\,\,\,\,\,\,\,\,\,\,\,\,\,\,
\,\,\,\,\,\,\,\,\,\,\,\,\,\,\,\,\,\,\,\,\,=\frac{n\kappa_{6}^{2}A\mathcal{T}^{t}_{r}}{4\alpha
H}\!+\!\frac{\kappa_{6}^{2}\mathcal{T}^{r}_{r}\!-\!\Lambda_{6}}{4\alpha}\!-\!\frac{3}{16\alpha^{2}}.
\label{ten}\end{eqnarray} For $\mathcal{T}^{t}_{r}=0$, equation
(\ref{viki}) is solved for $d\ln\!{\ss}/d\ln\!a$ as \bea&&
\frac{d\ln\!{\ss}}{d\ln\!a}=\frac{{\ss}P\!-\!6fA^{2}\!+\!1/2\alpha}{6(x\!-\!ka^{-2}\!-\!1/12\alpha)}.
\label{pasxa}\eea Differentiating equation (\ref{pasxa}) and
replacing in equation (\ref{ten}), we obtain the following algebraic
equation \bea
&&\!\!\!\!\!\!\!\!\!\!\!\!\!\!\!\chi_{5}\mathcal{A}^{5}\!+\!\chi_{4}\mathcal{A}^{4}\!+\!
\chi_{3}\mathcal{A}^{3}\!+\!\chi_{2}\mathcal{A}^{2}\!+\!\chi_{1}\mathcal{A}\!+\!\chi_{0}\!=\!0,
\label{arnio}\eea where $\mathcal{A}=A^{2}$, and $\chi$'s are
functions of $x$, $\varrho\!=\!-\sigma_{3}\rho$ given in the
appendix. Now, the system of equations (\ref{mammut}), (\ref{hope})
has been substituted equivalently by the system of equations
(\ref{pasxa}), (\ref{arnio}). Dropping from now on
$\mathcal{T}^{r}_{r}$ completely from the notation, differentiating
equation (\ref{arnio}) once more, and comparing with equation
(\ref{pasxa}), we finally substitute the system of equations
(\ref{mammut}), (\ref{hope}) by the algebraic system (\ref{arnio}),
(\ref{avga}): \bea &&\!\!\!\!\!\!\!
\psi_{7}\mathcal{A}^{7}\!\!+\!\psi_{6}\mathcal{A}^{6}\!+\!
\psi_{5}\mathcal{A}^{5}\!\!+\!\psi_{4}\mathcal{A}^{4}\!+\!
\psi_{3}\mathcal{A}^{3}\!\!+\!\psi_{2}\mathcal{A}^{2}\!\!+\!\psi_{1}\mathcal{A}
\!+\! \psi_{0}\!=\!0,\nn\\
&&\,\,\,\,\,\,\,\,\,\,\,\,\,\,\,\,\,\,\,\,\,\,\,\,\,\,\,\,\,\,\,\,\,\,\,\,\,\,\,\,\,\,\,\,\,\,\,\,\,\,\,\,\,\,
\label{avga} \eea where $\psi$'s are functions of $x, \varrho$,
given in the appendix. After some algebraic manipulation, the system
of equations (\ref{arnio}), (\ref{avga}) is written equivalently as
the following
system \bea &&\textrm{H}_{2}\mathcal{A}^{2}+\textrm{H}_{1}\mathcal{A}+1=0\label{magiritsa1}\\
&&\textsf{H}_{1}\mathcal{A}+\textsf{H}_{0}=0, \label{magiritsa2}
\eea where $\textrm{H}$'s, $\textsf{H}$'s are functions of $x,
\varrho$ given in the appendix. From equations (\ref{magiritsa1}),
(\ref{magiritsa2}) one obtains \bea \mathcal{H}(x,\varrho)\equiv
\textrm{H}_{2}\textsf{H}_{0}^{2}-\textrm{H}_{1}\textsf{H}_{0}
\textsf{H}_{1}+\textsf{H}_{1}^{2}=0. \label{antegeia} \eea This
equation could still be the (first order) Hubble equation of the
model even without containing any integration constants. However,
this is not the case, since the consistency of equation
(\ref{antegeia}) with equation (\ref{magiritsa2}) gives \bea
&&\!\!\!\!\!\!\!\!
\mathcal{J}\!(x,\varrho)\equiv\{3(1\!+\!w\!)x\varrho
\textsf{H}_{1}\!+\![(1\!+\!9w\!)\varrho\!-\!8(\!\sigma\!-\!3\sigma_{1}x\!)]
\textsf{H}_{0}\}\mathcal{H}_{,x}\nn\\
&&\,\,\,\,\,\,\,\,\,\,\,\,\,\,\,\,\,\,\,\,\,\,\,\,\,\,\,\,\,\,\,\,\,\,
+3(1\!+\!w\!)\varrho[(\varrho\!+\!\sigma\!)\textsf{H}_{1}\!+\!9\sigma_{1}\!\textsf{H}_{0}]
\mathcal{H}_{,\varrho}\!=\!0, \label{kalokairi}\eea where
$\sigma=\sigma_{2}+\sigma_{1}/4\alpha$. It can now be checked (e.g.
numerically) that on the two-dimensional plane $(x,\varrho)$ the two
curves $\mathcal{H}(x,\varrho)\!=\!0$,
$\mathcal{J}\!(x,\varrho)\!=\!0$ do not coincide, which completes
our statement of non-existence of isotropic braneworld cosmologies
\footnote {Attempting to generalize metric (\ref{cosmobulk}) to an
off-diagonal ansatz by adding the term
$2J^{i}(t,r)\gamma_{ij}(x)dtdx^{j}$, the matching conditions
(\ref{matching}) provide $J^{i} {}^{^{\prime}}(t,0^{+})\!=\!0$.
Additionally, in order to assure an untilded isotropic 4-cosmology,
one has to impose $J^{i}(t,0)\!=\!0$. Therefore, all the first order
brane equations implying the inconsistency remain unaffected.}.

If we are interested in looking at the compatibility of embedding a
maximally symmetric 3-brane (with $R\!=\!4\ell$) carrying only a
tension in a static bulk, we have to put in the line-element
(\ref{cosmobulk}) $L(t,r)\!=\!\tilde{L}(r)$ (thus $\beta$=constant),
$n(t,r)\!=\!\tilde{n}(r)$, and
$a(t,r)\!=\!\tilde{n}(r)\tilde{a}(t)$, where
$\dot{\tilde{a}}^{2}\!+\!k\!=\!\ell\tilde{n}(0)^{2}\tilde{a}^{2}/3$.
For the regular case (b), equations (\ref{roa}), (\ref{rua})
coincide giving
\bea\sigma_{2}\!+\!\beta/4\alpha\!=\!\ell(\sigma_{1}\!-\!\beta),\eea
equations (\ref{es}), (\ref{mammut}) are trivially satisfied, and
equation (\ref{hope}) gives the value of the bulk cosmological
constant \bea \Lambda_{6}\!=\!2\ell (1\!+\!2\alpha\ell/3), \eea
making the embedding of maximally symmetric branes permissible. This
solution generalizes known results from cosmic strings. For the case
(a), equation (\ref{es}) gives
\bea\sigma_{2}\!=\!\ell\sigma_{1},\eea the matching conditions
(\ref{roa}), (\ref{rua}) coincide giving \bea
A^{2}\!=\!N^{2}\!=\!(\ell\!+\!1/4\alpha)/3, \eea equation
(\ref{mammut}) is trivially satisfied, and equation (\ref{hope})
gives again a value for the bulk cosmological constant
\bea\Lambda_{6}\!=\!-5/12\alpha,\eea with the deficit angle $\beta$
remaining undetermined. This is a new solution with a maximally
symmetric 3-brane embedded in a six-dimensional bulk with negative
cosmological constant (non-$AdS_{6}$), where divergences of the bulk
scalar curvature of the form $1/r$ appear as approaching the brane.

In conclusion, we have considered a codimension two (thin)
braneworld with conical singularities in Einstein-Gauss-Bonnet
(-induced gravity) theory with a bulk cosmological constant, where
the addition of the Gauss-Bonnet term is known to make meaningful
the situation when non-trivial braneworld matter content is
included. Considering all the field equations at the position of the
brane, we have shown that for axially symmetric bulks an isotropic
braneworld cosmological ansatz is incompatible with the model.
Technically, this is because there is (excluding the gauge
arbitrariness) one equation more than the unknowns, which is finally
inconsistent with the other equations. Having developed to some
degree our formulation on a general basis, makes it also applicable
to other braneworld configurations. It is easily seen that the case
of a maximally symmetric 3-brane is compatible with the formulation.

\[ \]
{\bf Acknowlegements} We wish to thank C. Carvalho, C. Charmousis,
T. Christodoulakis, R. Emparan, K. Koyama, R. Maartens, E.
Papantonopoulos, E. Verdaguer and in particular J. Garriga for
useful discussions. This work is supported by a European Commission
Marie Curie Fellowship, under contract MEIF-CT-2004-501432.

\[ \]
{\bf Appendix} We provide here the quantities $\chi(x,\varrho)$
appearing in equation (\ref{arnio})
\begin{eqnarray}
&&\!\!\!\!\!\!\!\chi_{5}\!=\!9\sigma_{1}^{2}\{(1\!+\!9w\!)\varrho\!-\!4[2\sigma\!-\!9\sigma_{1}(x\!-\!
\tilde{\alpha}\!-\!\tilde{k})]\}\nn
\end{eqnarray}
\begin{eqnarray}
&&\!\!\!\!\!\!\!\frac{\chi_{4}}{3\sigma_{\!1}}\!=\!-2(1\!+\!12w\!+\!27w^{\!2})\varrho^{\!2}\!\!+\!
\{4\sigma(5\!+\!21w\!)\!-\!3\sigma_{\!1}[12\tilde{\alpha}(27w^{\!2}\nn\\
&&\,\,\,\!+30w\!+\!2)\!+\!18\tilde{k}(1\!+\!17w\!+\!18w^{\!2})\!-\!(13\!+\!261w\!+\!324w^{\!2})\nn\\
&&\,\,\,\,x]\}\varrho\!-\!4[8\sigma^{2}\!\!-\!3\sigma\sigma_{\!1}(19x\!+\!3\tilde{\alpha}\!-\!9\tilde{k})
\!-\!54\sigma_{\!1}^{2}x(x\!-\!\tilde{\alpha}\!-\!\tilde{k}\!)]\nn
\end{eqnarray}
\begin{eqnarray}
&&\!\!\!\!\!\!\!\chi_{3}\!=\!(1\!+\!15w\!+\!54w^{\!2})\varrho^{3}\!\!-\!6\{\!(2\!+\!13w\!-\!9w^{\!2})
\sigma\!+\!2\sigma_{\!1}\![(2\!+\!30w\nn\\
&&\!\!\!+27w^{\!2})x\!-\!3\tilde{a}(2\!+\!23w\!+\!18w^{\!2})\!-\!\tilde{k}(4\!+\!51w\!+\!54w^{\!2})]\}
\varrho^{\!2}\!\!+\!3\nn\\
&&\!\!\!\{\!(9\!-\!31w\!)\sigma^{\!2}\!\!-\!4\sigma\sigma_{\!1}\![3\tilde{\alpha}(1\!-\!\!23w\!-\!\!18w^{\!2})\!+
\!2(27w^{\!2}\!\!+\!42w\!+\!\!14)\nn\\
&&\!\!\!x\!-\!\tilde{k}(11\!\!+\!51w\!+\!54w^{\!2}\!)]\!+\!\!3\sigma_{\!1}^{\!2}x[36\tilde{\alpha}(2\!+\!
13w\!+\!9w^{\!2}\!)\!-\!(324w^{\!2}\nn\\
&&\!\!\!+297w\!+\!53)x\!+\!\!12\tilde{k}(5\!+\!30w\!+\!27w^{\!2}\!)]\}\varrho\!+\!
2[20\sigma^{\!3}\!\!+\!\!243\sigma_{\!1}^{\!3}(\!3\omega\nn\\
&&\!\!\!-\!2x^{\!2})(\!x\!-\!\tilde{\alpha}\!-\!\tilde{k}\!)
\!+\!6\sigma^{\!2}\sigma_{\!1}\!(\!x\!\!-\!9\tilde{\alpha}\!+\!\!7\tilde{k}\!)\!-\!
36\sigma\sigma_{\!1}^{\!2}x(\!10x\!+\!9\tilde{\alpha}\!-\!\!3\tilde{k}\!)]\nn
\end{eqnarray}
\begin{eqnarray}
&&\!\!\!\!\!\!\!\chi_{2}\!=\!27\sigma_{\!1}^{\!2}x^{\!3}[(13\!+\!9w)\varrho\!+\!4\sigma]\!-\!\!2\tilde{\alpha}
(\varrho\!+\!\sigma\!)[2\sigma(9w^{\!2}\!\!+\!45w\!+\!2)\varrho\nn\\
&&-34\sigma^{\!2}\!-\!243\sigma_{\!1}^{\!2}\omega\!+\!2(1\!+\!9w\!-\!9w^{\!2})\varrho^{\!2}]\!+\!6\sigma_{\!1}
x^{\!2}\{\!(81w^{\!2}\!\!+\!72w\nn\\
&&+11)\varrho^{\!2}\!\!+\!10\sigma(2\sigma\!+\!9\tilde{\alpha}\sigma_{\!1}\!)\!+\!2\varrho[(29\!+\!63w
\!+\!54w^{\!2})\sigma\!-\!9\tilde{\alpha}\sigma_{\!1}\nn\\
&&(4\!+\!9w\!)]\}\!-\!x(\varrho\!+\!\sigma\!)\{\!(1\!+\!21w\!+\!144w^{\!2})\varrho^{\!2}
\!\!+\!124\sigma^{\!2}\!\!+\!486\sigma_{\!1}^{\!2}\omega
\nn\\
&&\!-360\tilde{\alpha}\sigma\sigma_{\!1}\!-\![(55\!+\!339w\!+\!36w^{\!2})\sigma\!-\!\!
72\tilde{\alpha}\sigma_{\!1}(2\!+\!\!16w\!+\!9w^{\!2})]\nn\\
&&\varrho\}\!-\!2\tilde{k}\{\!(1\!+\!9w\!-\!18w^{\!2})\varrho^{\!3}\!\!+\!6[(2\!+\!15w)\sigma
\!+\!2\sigma_{\!1}(27w^{\!2}\!\!+\!30w\nn\\
&&+4)x]\varrho^{2}\!\!-\!\sigma[26\sigma^{\!2}\!\!-\!12\sigma\sigma_{\!1}x\!+\!27\sigma_{\!1}^{\!2}
(2x^{\!2}\!+\!9\omega\!)]\}\!-\!6\tilde{k}\varrho\{4\sigma\sigma_{\!1}\nn\\
&&\!(5\!+\!\!30w\!+\!\!27w^{\!2})x\!+\!\!9\sigma_{\!1}^{\!2}[(7\!\!+\!9w\!)x^{\!2}\!\!-\!9\omega\!]\!-\!
\sigma^{\!2}(5\!-\!27w\!-\!6w^{\!2})\!\}\nn
\end{eqnarray}
\begin{eqnarray}
&&\!\!\!\!\!\!\!\frac{\chi_{1}}{\varrho\!+\!\sigma}\!=\!
2\tilde{k}\{2x[(1\!+\!6w\!-\!9w^{\!2})\varrho^{\!2}\!+\!(5\!+\!42w\!+\!9w^{\!2})\sigma\!\varrho
\!-\!14\sigma^{\!2}]\nn\\
&&+6\sigma_{\!1}x^{\!2}[(4\!+\!9w\!)\varrho\!-\!5\sigma]\!-\!\!27\sigma_{1}\omega(\varrho\!+\!\sigma\!)\}
\!-\!18\sigma_{\!1}\!(\varrho\!+\!\sigma\!)(3\tilde{\alpha}\omega\nn\\
&&\!+2x^{\!3})\!+\!2x\{2\tilde{\alpha}(2\!+\!15w\!-\!9w^{\!2})\varrho^{\!2}
\!\!+\![27\sigma_{\!1}\omega\!-\!\!2\tilde{\alpha}\sigma(2\!-\!51w\nn\\
&&-9w^{\!2})]\varrho\!+\!\sigma(27\sigma_{\!1}\omega\!-\!44\tilde{\alpha}\sigma\!)\}\!-\!x^{\!2}
\{\!(1\!-\!9w\!-\!90w^{\!2})\varrho^{\!2}\!\!-4\sigma\nn\\
&&(20\sigma\!-\!63\tilde{\alpha}\sigma_{\!1}\!)\!-\!\![36\tilde{\alpha}\sigma_{\!1}\!(2\!+\!9w\!)
\!-\!\!(47\!+\!243w\!+\!36w^{\!2})\sigma]\varrho\}\nn
\end{eqnarray}
\begin{eqnarray}
&&\!\!\!\!\!\!\!\frac{\chi_{0}}{(\varrho\!+\!\sigma\!)^{2}}\!=\!x^{2}\{[(1\!-\!3w)\varrho\!+\!4\sigma]x
\!-\!4\tilde{\alpha}[(1\!+\!6w)\varrho\!-\!5\sigma]\nn\\
&&\,\,\,\,\,\,\,\,\,\,\,\,\,\,\,\,\,\,\,\,\,\,\,\,\,\,
-2\tilde{k}[(1\!+\!3w)\varrho\!-\!2\sigma]\}\!-\!2\omega(\varrho\!+\!\sigma\!)
(x\!-\!\tilde{\alpha}\!-\!\tilde{k}),\nn\label{ole}
\end{eqnarray}
where $\tilde{\alpha}\equiv 1/12\alpha$,
$\omega\equiv(\Lambda_{6}\!-\!\kappa_{6}^{2}\mathcal{T}^{r}_{r}\!+\!5/12\alpha)/6\alpha$,
and $\tilde{k}\equiv k/a^{2}=k(\varrho/\varrho_{o})^{2/3(1+w)}$,
with $\varrho_{o}\!>\!0$ integration constant.

We give here the quantities $\psi(x,\varrho)$ appearing in equation
(\ref{avga})
\begin{eqnarray}
&&\!\!\!\!\!\!\!\psi_{j}\!=\!\tilde{\psi}_{j}\!+\!c_{j}\,\,\,\,\,,\,\,\,\,\,0\leq
j\leq 7,\nn\end{eqnarray} where
\begin{eqnarray}
&&\!\!\!\!\!\!\!\tilde{\psi}_{j}\!=\!\!=\!2(x\!-\!\tilde{\alpha}\!-\!\tilde{k})\{3(1\!+\!w)\varrho[x\chi_{j,x}
\!+\!(\varrho\!+\!\sigma\!)\chi_{j,\varrho}]\!-\![(1\!+\!9w\!)\varrho\nn\\
&&\,\,\,\,\,\,\,\,\,\,\,\,\,\,\,\,\,\,\,\,\,\,\,\,\,\,\,\,\,\,\,\,\,\,\,\,\,
-8(\!\sigma\!\!-\!\!3\sigma_{1}x\!)]\chi_{j-1,x}\!-\!27(1\!+\!w\!)\sigma_{1}\varrho\chi_{j-1,\varrho}\}
\nn\end{eqnarray}
\begin{eqnarray}
&&\!\!\!\!\!\!\!(c_{7},c_{6},c_{5},c_{4},c_{3},c_{2},c_{1},c_{0})
\!=\!(-15\sigma_{1}\chi_{5}\,,\,-12\sigma_{1}\chi_{4}\!+\!5\hat{\zeta}\chi_{5},\nn\\
&&-9\sigma_{1}\chi_{3}\!+\!4\hat{\zeta}\chi_{4}\!-\!5\check{\zeta}\chi_{5}\,,\,
-6\sigma_{1}\chi_{2}\!+\!3\hat{\zeta}\chi_{3}\!-\!4\check{\zeta}\chi_{4}\!+\!5\zeta\chi_{5},\nn\\
&&-3\sigma_{1}\chi_{1}\!+\!2\hat{\zeta}\chi_{2}\!-\!3\check{\zeta}\chi_{3}\!+\!4\zeta\chi_{4}\,,\,
\hat{\zeta}\chi_{1}\!-\!2\check{\zeta}\chi_{2}\!+\!3\zeta\chi_{3},\nn\\
&&-\check{\zeta}\chi_{1}\!+\!2\zeta\chi_{2}\,,\,\zeta\chi_{1})\nn\end{eqnarray}
\begin{eqnarray}
&&\!\!\!\!\!\!\!\zeta\!=\!x(x\!+\!2\tilde{\alpha})(\varrho\!+\!\sigma\!)
\nn\end{eqnarray}
\begin{eqnarray}
&&\!\!\!\!\!\!\!
\hat{\zeta}\!=\!(1\!+\!6w\!)\varrho\!-\!5\sigma\!-\!6\sigma_{1}\!(6x\!-\!11\tilde{\alpha}\!-\!8\tilde{k})
\nn\end{eqnarray}
\begin{eqnarray}
&&\!\!
\check{\zeta}\!=\!6\tilde{\alpha}[(1\!-\!2w\!)\varrho\!+\!3\sigma]\!+\!4\tilde{k}[(1\!-\!3w\!)\varrho\!+\!4\sigma]
\!+\!9\sigma_{1}x^{\!2}\nn\\
&&\,\,\,\,\,\,\,-2x[(1\!-\!9w\!)\varrho\!+\!10\sigma\!-\!9\tilde{\alpha}\sigma_{1}].
\nn\end{eqnarray}

We provide now the quantities $\textrm{H}(x,\varrho)$,
$\textsf{H}(x,\varrho)$ appearing in equations (\ref{magiritsa1}),
(\ref{magiritsa2})
\begin{eqnarray}
&&\!\!\!\!\!\!\!(\textrm{H}_{2},\textrm{H}_{1})\!=\!\Big(\!\!F_{3}[(\!C_{2}\!-\!F_{1}C_{1}\!)
(\!B_{1}\!\!-\!p_{1}\!)\!+\!F_{1}\!(\!B_{2}\!-\!p_{2}\!)\!+\!p_{3}\!-\!B_{3}],\nn\\
&&(F_{1}F_{2}\!-\!F_{3})
[B_{2}\!-\!p_{2}\!-\!C_{1}\!(\!B_{1}\!\!-\!p_{1}\!)]\!-\!(\!C_{4}-\!F_{2}C_{2}\!)(\!B_{1}\!\!-\!p_{1}\!)\nn\\
&&-F_{2}(\!B_{3}\!-\!p_{3}\!)\!+\!B_{5}\!-\!p_{5}\!\!\Big)\Big{/}\Big[\!(F_{1}^{2}\!-\!F_{2})[B_{2}\!-\!p_{2}\!-\!C_{1}\!
(\!B_{1}\!\!-\!p_{1}\!)]\nn\\
&&-(C_{3}\!-\!F_{1}C_{2}\!)(\!B_{1}\!\!-\!p_{1}\!)\!-\!F_{1}\!(\!B_{3}\!-\!p_{3}\!)\!+\!B_{4}\!-\!p_{4}\!\Big]\nn
\end{eqnarray}
\begin{eqnarray}
&&\!\!\!\!\!\!\!(\textsf{H}_{1},\textsf{H}_{0}\!)\!=\!\Big(\!\!
(F_{3}\!-\!\textrm{H}_{2}F_{1})[B_{2}\!-\!p_{2}\!-\!C_{1}(\!B_{1}\!-\!p_{1}\!)]\!+\!(C_{4}\!\!-\!\textrm{H}_{2}C_{2})\nn\\
&&(\!B_{1}\!\!-\!p_{1}\!)\!+\!\textrm{H}_{2}(\!B_{3}\!-\!p_{3}\!)\!+\!p_{5}\!-\!B_{5},
(F_{2}\!-\!\textrm{H}_{1}F_{1}\!)[B_{2}\!-\!p_{2}\!-\!C_{1}\nn\\
&&(\!B_{1}\!\!-\!p_{1}\!)]\!+\!(C_{3}\!-\!\textrm{H}_{1}C_{2})(\!B_{1}\!\!-\!p_{1}\!)\!+\!\textrm{H}_{1}
(\!B_{3}\!-\!p_{3})\!+\!p_{4}\!-\!B_{4} \!\Big)\nn
\end{eqnarray}
where
\begin{eqnarray}
&&\!\!\!\!\!\!\!(F_{3},F_{2},F_{1}\!)\!=\!\Big(\!\!C_{4}[C_{1}\!(\!B_{1}\!-\!p_{1}\!)\!+\!p_{2}\!-\!B_{2}],
(\!C_{1}\!C_{3}\!-\!C_{4}\!)(\!B_{1}\!\!-\!p_{1}\!)\nn\\
&&-C_{3}\!(B_{2}\!-\!p_{2}\!)\!+\!B_{5}\!-\!p_{5},(\!C_{1}\!C_{2}\!-\!C_{3}\!)(\!B_{1}\!-\!p_{1}\!)\!-\!
C_{2}(B_{2}\!-\!\!p_{2}\!)\nn\\
&&+B_{4}\!-\!p_{4}\!\!\Big)/\,[(C_{1}^{2}\!-\!C_{2})(\!B_{1}\!-\!p_{1}\!)\!-\!C_{1}\!
(B_{2}\!-\!p_{2}\!)\!+\!B_{3}\!-\!p_{3}]\nn
\end{eqnarray}
\begin{eqnarray}
&&\!\!\!\!\!\!\!(C_{4},C_{3},C_{2},C_{1}\!)\!=\!\Big(\!\!B_{5}(\!B_{1}\!\!-\!p_{1}\!),B_{4}\!(\!B_{1}\!\!-\!p_{1}\!)
\!+\!p_{5}\!-\!B_{5},B_{3}(\!B_{1}\nn\\
&&\!-p_{1}\!)\!+\!p_{4}\!-\!B_{4},B_{2}(\!B_{1}\!\!-\!p_{1}\!)\!+\!p_{3}\!-\!\!B_{3}\!\!\Big)/(\!B_{1}^{2}\!\!-\!B_{2}\!-
\!\!B_{1}p_{1}\!\!+\!p_{2}\!)\nn
\end{eqnarray}
\begin{eqnarray}
&&\!\!\!\!\!\!\!(\!B_{5},B_{4},B_{3},B_{2},B_{1}\!)\!=\!\Big(\!\!q_{7},p_{5}\!(p_{1}\!\!-\!q_{1}\!)\!+\!q_{6},p_{4}\!
(p_{1}\!\!-\!q_{1}\!)\!+\!q_{5}\!\!-\!p_{5},\nn\\
&&\!\!p_{3}\!(p_{1}\!\!-\!q_{1}\!)\!+\!q_{4}\!\!-\!p_{4},p_{2}(p_{1}\!\!-\!q_{1}\!)\!+\!q_{3}\!-\!p_{3}\!\!\Big)/
(p_{1}^{2}\!\!-\!p_{2}\!-\!p_{1}q_{1}\!\!+\!q_{2}\!)\nn
\end{eqnarray}
and $p_{i}=\chi_{i}/\chi_{0}$ ($1\leq i \leq 5$), $q_{j}=
\psi_{j}/\psi_{0}$ ($1\leq j \leq 7$).

\end{document}